\documentclass{article}

\usepackage{PRIMEarxiv}

\usepackage[utf8]{inputenc} 
\usepackage[T1]{fontenc}    
\usepackage{hyperref}       
\usepackage{url}            
\usepackage{booktabs}       
\usepackage{amsfonts}       
\usepackage{nicefrac}       
\usepackage{microtype}      
\usepackage{fancyhdr}       
\usepackage{graphicx}
\usepackage{amsmath}
\usepackage{amsbsy}
\usepackage{mathtools}
\graphicspath{{media/}}     

\title{Optimal Snake Locomotion on Flat Surfaces: An Analytical Framework
}

\author{
  Yogesh Phalak, Hodjat Pendar \\
  Department of Mechanical Engineering \\
  Virginia Tech \\
  Blacksburg\\
  \texttt{\{Hodjat Pendar\}hpendar@vt.edu} \\
}

\begin{document}
\maketitle

\begin{abstract}
In this theoretical study, we present an analytical framework to investigate the slithering motion of snakes on flat surfaces. While previous studies have predominantly relied on numerical methods to identify optimal locomotion kinematics, such approaches are often sensitive to initial guesses and the number of kinematic parameters in the model. Here, we derive analytical solutions for optimal kinematics that minimize the cost of transport or maximize the velocity under varying friction anisotropy conditions. Our analysis assumes a uniform weight distribution and negligible body rigidity, though the framework can be extended to more complex scenarios. Furthermore, we demonstrate the applicability of this approach to the undulatory motion of other elongated bodies in various media, where interactive forces can be described using resistive force theory, such as swimming through sand or viscous fluids.
\end{abstract}

\keywords{Snake locomotion, optimization, friction anisotropy, resistive force theory}

\section{Introduction}
Snake locomotion on flat surfaces critically depends on the frictional anisotropy of their ventral scales \cite{doi:10.1073/pnas.0812533106,hu2012slithering}.
Unlike most terrestrial animals, which rely on static friction between their appendages and the ground to move, snakes mostly utilize kinetic friction to move while simultaneously expending energy to slide their bodies along substrates. The friction force functions both as a propulsive and a resistive force simultaneously. Consequently, they have adapted mechanisms to perform locomotion efficiently. The evolutionary adaptation of snake scales is a key factor in their ability to move effectively through different environments. These scales have specialized microstructures that create anisotropic friction, meaning the frictional resistance varies depending on the direction of movement. This allows snakes to minimize resistance when moving forward while providing enough grip to propel themselves \cite{10.1242/jeb.23.2.101, https://doi.org/10.1111/nyas.15109}.
In 2009 David Hu and his colleagues experimentally demonstrated that friction anisotropy is necessary for slithering on flat surfaces. By sliding 10 species of milk and corn snakes in 9 sliding directions on cloth and fiberboard on an incline, they showed how friction coefficient varies depending on the direction of slide. They demonstrated that the friction coefficient in the longitudinal and normal directions at any point of the body depends on the angle between the velocity of that point and the tangential direction of the body. The friction anisotropy has been examined in other species and it is shown that the friction is different in cranial, caudal, and lateral directions \cite{doi:10.1073/pnas.0812533106, doi:10.1137/1.9781611970517}.

Snakes employ various locomotion patterns to navigate diverse substrates, with the kinematics of their undulatory movements playing a crucial role in determining locomotion efficiency and performance \cite{https://doi.org/10.1111/nyas.15109, a9685a0d-387e-386e-9f8f-0aa41d2ab5f6}. Several studies have investigated the optimal locomotion patterns of snakes on flat surfaces \cite{doi:10.1098/rspa.2013.0236, hu2012slithering, C7SM01545C}. Silas Alben developed a numerical method to determine the optimal motion of snakes on flat surfaces across a wide range of frictional anisotropy \cite{doi:10.1098/rspa.2013.0236}. He approximated the periodic deformation of the snake's body over one cycle using a double series expansion with Chebyshev polynomials. For given sets of frictional parameters, the expansion coefficients were computationally determined to maximize efficiency.

These computational studies provide valuable insights into how body deformation parameters affect locomotion efficiency across different substrates. However, establishing a fundamental theoretical framework can help us better identify the key components of undulation and determine the maximum possible efficiency of snake locomotion. In this work, we provide an analytical framework for theoretically studying snake locomotion and to determine the effect of each principal component of kinematics to move on different substrates with various degrees of friction anisotropy.  We use this framework to find patterns that minimize energy consumption and also patterns that maximize locomotion speed for elongated bodies.


The frictional force is the dominant factor in snake dynamics, while inertial forces are negligible in comparison \cite{doi:10.1073/pnas.0812533106, doi:10.1098/rspa.2022.0312, doi:10.1098/rspa.2013.0236, PhysRevE.89.012717}. This characteristic makes snake locomotion similar to movement in viscous fluids, where viscous forces dominate. The other similarity between the physics of snake locomotion and motion in viscus flow is that the interactive forces can be described using the Resistive Force Theory (RFT). The RFT developed by Gray and Hancock 1955 \cite{10.1242/jeb.32.4.802}, further used by Lighthill 1975 \cite{doi:10.1137/1.9781611970517} to study flagellar propulsion in fluids has since been widely applied to analyze the motion of slender bodies in viscous fluids. In this method, a body is partitioned into infinitesimal segments. At low Rwynolds number the drag forces in tangential and normal directions at each segment are proportional to the velocity of the fluid with respect to the segment in the tangential and normal directions. RFT is a powerful method to analyze the locomotion of microorganisms and objects in viscous fluids \cite{10.1119/1.10903, 10.1039/9781782628491-00100, C6SM01636G, PhysRevFluids.1.053202}. In the past decade the traditional RFT framework, which was originally designed for viscous fluids, has been adapted to model the interaction of objects with a granular media  \cite{Maladen2009, doi:10.1126/science.1229163, Aguilar2015RobophysicalSO, 10.1063/1.4898629}. The extended RFT model, verified through extensive experimentation, is used to understand and model how organisms and robots move through environments composed of discrete particles. In the RFT for granular media, the tangential and normal forces at each segment of the body depend on the orientation of the local velocity vector relative to the segment's direction. However, unlike in viscous fluids, the forces are not directly proportional to the velocity itself.David Hu et al. \cite{doi:10.1137/1.9781611970517} also demonstrated that the frictional forces between snakes and various substrates follow the same pattern, with the friction coefficients in the tangential and normal directions being proportional to the cosine and sine of the angle between the velocity of a point and the tangential axes at that point.

Recognizing the parallels between the locomotion of microorganisms in viscous fluids and the friction-based locomotion of snakes, we extend Lighthill's well-established method for determining the optimal locomotion of spermatozoa \cite{doi:10.1137/1.9781611970517} to identify the optimal undulation patterns for snakes across various substrates. Lighthill simplifies the movement equations of a flagellum by assuming periodic motion of the flagellum and steady and linear movement of the microorganism. He then analytically and numerically maximizes propulsion efficiency, defined as the ratio of forward thrust generated to the energy expended in moving the flagellum. Despite the fundamental differences in the physics of snake locomotion and microorganism movement, we extend this approach to identify the optimal patterns of undulation in snake locomotion. We simplify the complex time-dependent dynamical equations of locomotion into a time-independent form, facilitating analytical analysis. We then determine the optimal shape for both maximum speed and minimal cost of transport. These findings offer valuable insights into understanding the diverse patterns of snake locomotion. The results of this work can also be applied to the development and improvement of snake robots.

\section{Model}

\begin{figure}[ht]
  \centering
  \includegraphics[width=\textwidth]{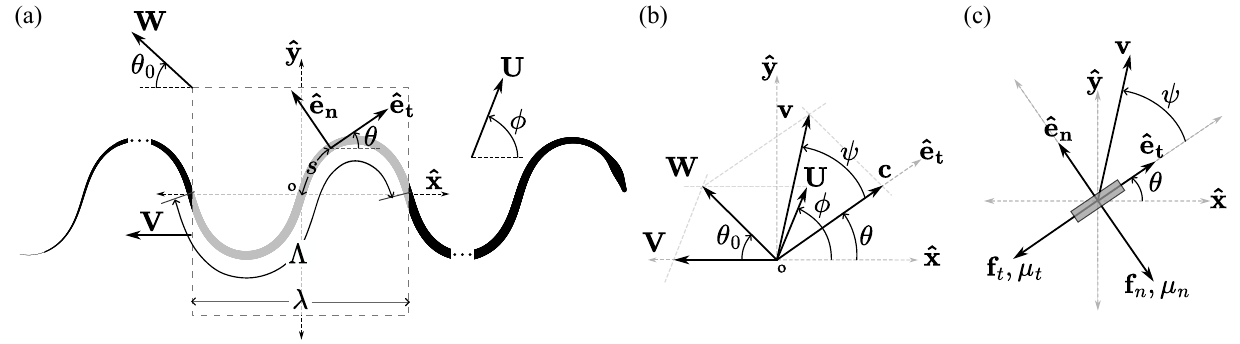}
  \caption{Kinematic parameters of snake locomotion: (a) schematics of snake locomotion, (b) Velocity vectors schematic,(c) The friction components that can be determined using resistive force theory.}
    \label{fig:RFTSchematic}
\end{figure}

In this model, we assume that the snake's body is uniform, inextensible, flexible, and lacks rigidity, moving in a 2D horizontal plane. Following Lighthill's work on flagella, we assume the snake's body shape is a periodic function along axis x and that the snake moves at a steady speed $U$ with an angle $\phi$ relative to the direction of the snake's body (Figure \ref{fig:RFTSchematic}). In a reference frame moving with the propulsive wave, the snake has only a constant tangential velocity along the body waveform (Figure \ref{fig:RFTSchematic}), which is considered to be constant $c$. In this frame, the snake's body shape is denoted by the slope angle function $\theta(s)$, which is a continuous and periodic function. Here, $s$ represents the arc length measured from the middle of a cycle (Figure \ref{fig:RFTSchematic}). 

The propulsive wave propagates along the body from the head to the tail with the velocity of $\boldsymbol{V} = -\lambda \hat{i}$, where $\lambda$ is the wavelength. The frame that moves along with a propulsive wave has a velocity of $\boldsymbol{W} = \boldsymbol{U}+\boldsymbol{V}$ with respect to the ground. In this frame, the trace of the movement appears to be fixed, and all the points on the body flow along that trace with a velocity of $\boldsymbol{c} = c \, \hat{\boldsymbol{e}}_t$. This speed can be determined as $c =\frac{\Lambda}{T}$, where $\Lambda$ and $T$ are the arc length of one cycle and the period respectively. The absolute velocity of any arbitrary point on the body, as illustrated in Figure \ref{fig:RFTSchematic}(b), is $\boldsymbol{v}(s) = \boldsymbol{W}+\boldsymbol{c}(s)$.

To non-dimensionalize the quantities based on the characteristic parameters we non-dimensionalized the length by the arc length of one cycle ($\Lambda$), time by the period of undulation ($T$), and the mass by $\rho \Lambda$, where $\rho$ is the density per length. After non-dimensionalion $c$ will be equal to 1 and the velocity of the body points is $\boldsymbol{v} = \boldsymbol{W} + \boldsymbol{c} = \boldsymbol{U} - \lambda\hat{\boldsymbol{i}} + \hat{\boldsymbol{e}}_t$, where $\lambda = \int_{-0.5}^{0.5} cos(\theta) ds$ is the wavelength. Here, $\theta$ is the angle between $\hat{\boldsymbol{e}}_t$ and $x$ axis. The magnitude of the velocity in an arbitrary point can be determined as (Figure \ref{fig:RFTSchematic} (b))   

\begin{equation}
v(s) = \left( 1+W^2-2 W \cos{(\theta+\theta_0)} \right)^{1/2} = \cos{\psi}-\left(W^2-\sin^2{\psi}\right)^{1/2}   \label{eq:LocalVel}  
\end{equation}
where $\theta_0$ is the angle between $\boldsymbol{W}$ and $-x$ axis (Figure \ref{fig:RFTSchematic} (b)), and $\psi(s)$ is the angle between the velocity $\boldsymbol{v}$ of a body point and the tangential direction $\hat{\boldsymbol{e}}_t$. $\psi$ has an important role in the friction that we will discuss it later. There are relationships between $\psi$ and $\theta$, which we will use them later to simplify the equations. From Figure \ref{fig:RFTSchematic} (c) we have:

\begin{align}
    \sin\psi &= \frac{W}{v}\sin(\theta+\theta_0), \label{eq:psi_theta_1}\\
    \cos\psi &= \frac{1}{v}\left(1-W \cos(\theta+\theta_0)\right), \label{eq:psi_theta_2} 
\end{align}

The original investigation into the friction properties of snake scales was carried out by Hu and Shelley \cite{hu2012slithering}. They developed a mathematical model to illustrate the friction anisotropy between snakes and substrates (\ref{eq:AnisotropyModel}). This model, which closely aligns with observational data, has been widely embraced in the study of snake locomotion \cite{doi:10.1073/pnas.0812533106, hu2012slithering, doi:10.1098/rspa.2022.0312, doi:10.1098/rspa.2013.0236, PhysRevE.89.012717}. Based on this model the friction is characterized by three friction coefficients: $\mu_f$ (forward), $\mu_b$ (backward), and $\mu_n$ (normal), representing the frictional forces acting on the snake's body. The Coulomb friction force per unit length acting on any point on the body is:

\begin{align}
    \boldsymbol{F}(s,t) &= -\rho g \left(\mu_n \sin{\psi} \hat{\boldsymbol{e}}_n + \mu_t \cos{\psi}\hat{\boldsymbol{e}}_t\right), \label{eq:AnisotropyModel} 
\end{align}

where  $\mu_t = (\mu_f + \mu_b) H(\cos{\psi}) - \mu_b $, and $\rho$ is the mass density per length. In this equation, $H$ is a Heavyside function. The unit vectors $\hat{\boldsymbol{e}}_n$  and $\hat{\boldsymbol{e}}_t$ show the normal and tangential directions at the arbitrary point on the snake body.

After non-dimensionalizing the variables the dynamic equations based on Newton's second law were derived: 

\begin{align}
    \frac{\Lambda}{\mu_f g T^2}\int{\Ddot{\boldsymbol{x}}}ds &= \int{\boldsymbol{f}}ds, \label{eq:force_balance}\\
    \frac{\Lambda}{\mu_f g T^2} \int{\boldsymbol{x}\times\Ddot{\boldsymbol{x}}}ds &= \int{\boldsymbol{x}\times\boldsymbol{f}}ds, \label{eq:moment_balance} 
\end{align}

where $\boldsymbol{x}$ denotes the position of the body points in an inertial frame and $\boldsymbol{f}$ is the non-dimensionalize form of $\boldsymbol{F}$ where it is divided to $\mu_f$ as well. For most snakes $\frac{\Lambda}{\mu_f g T^2}<<1$ and can be approximated as zero \cite{doi:10.1073/pnas.0812533106, hu2012slithering}. This implies that inertia is negligible compared to frictional forces. This model has been widely used in various studies for snakes with arbitrary shapes that can also vary over time. Following Lighthill's work on flagella, we assume the snake's body consists of many periodic cycles, ideally infinite, allowing the snake to move steadily without rotation. If the snake moves at a constant speed, we can apply the aforementioned Newton's equations to one cycle of the snake's body in a reference frame moving with the propulsive wave. In this frame, the body shape remains constant, rendering the equations time-independent. When considering one cycle of the body, internal forces are included in Newton's equations. However, the internal forces at the two ends of each cycle are equal and opposite, thus canceling each other out in the equation \ref{eq:psi_theta_1}. Equation \ref{eq:psi_theta_2} can be used to determine the internal forces and moments, which we do not use it here for the locomotion analysis.

In this study, we initially assume that the body slides only in the forward direction. Later, we prove that with the assumptions of this model, all the points on the body can only slide forward or backward in addition to sliding in the lateral direction, i.e. $\psi<\pi/2$ for all $s$ or  $\psi>\pi/2$ for all the body points. This is typically the case when a snake continuously moves on a flat surface. Therefore the friction can be determined as:

\begin{equation}
\boldsymbol{f}(s) = -\left( \cos\psi \, \hat{\boldsymbol{e}}_t + \mu\sin\psi \, \hat{\boldsymbol{e}}_n \right)
\label{eq:friction}
\end{equation}

where $\mu = \mu_n/\mu_f$ is the friction coefficient ratio. The components of the friction in the x and y directions are $f_x(s) = -\left( \cos\psi \cos\theta - \mu\sin\psi \sin\theta \right)$ and $f_y(s) = -\left( \cos\psi \sin\theta + \mu\sin\psi \cos\theta \right)$.
Theoretically, a friction ratio of 1 corresponds to an isotropic substrate. According to multiple experimental studies \cite{doi:10.1073/pnas.0812533106, hu2012slithering, 10.1242/jeb.23.2.101, https://doi.org/10.1111/nyas.15109, doi:10.1137/1.9781611970517, a9685a0d-387e-386e-9f8f-0aa41d2ab5f6, doi:10.1098/rspa.2013.0236}, the friction ratio is typically greater than 1 and lies between 1 and 2. 
For further investigation, we will keep $\mu \geq 1$. The internal forces at two ends of one cycle are equal and opposite directions. Therefore the total friction force acting on the body within one cycle is:

\begin{equation}
\int_{-0.5}^{0.5}{\boldsymbol{f}(s) \, ds} = \boldsymbol{0},
\label{eq:NewtonsEquation}
\end{equation}

The average velocity of each cycle of the body is equal to the total velocity of the snake, i.e. $\boldsymbol{U} = \int_{-0.5}^{0.5}{\boldsymbol{v}(s) \, ds}$. Splitting this equation in the direction of the total velocity and perpendicular to it results in: 

\begin{align}
    U &= \int_{-0.5}^{0.5}{v\cos \left(\theta + \psi - \phi \right)}ds, \label{eq:U_equation}\\
    0 &= \int_{-0.5}^{0.5}{v\sin \left(\theta + \psi - \phi \right)}ds, 
\end{align}
where $\phi$ is the angle between $\boldsymbol{U}$ and $x$ axis. 

\section{Results}

In animal locomotion, the overall efficiency of movement is measured by a dimensionless quantity known as the metabolic cost of transport \cite{10.1242/jeb.163.1.1}. It is defined as the ratio of the power input required to move the body at a constant velocity to the product of the total weight and the magnitude of that constant velocity. In this model, we are neglecting the elastic energy of the body, so the power input is solely the work done by friction forces. Considering the weight scaling in the model, the cost of transport can be expressed as

\begin{equation}
    \eta^{-1} = \frac{-1}{U}\int_{-0.5}^{0.5}{\boldsymbol{f}\cdot\boldsymbol{v} \, ds}, \label{eq:GenCostOfTransport}
\end{equation}

This equation can be simplified more by adding $\frac{1}{U}\int_{-0.5}^{0.5}{\boldsymbol{f}\cdot\boldsymbol{W} \, ds}$ to the equation. Because $\boldsymbol{W}$ is a constant vector and also $\int_{-0.5}^{0.5} \boldsymbol{f}\,ds = \boldsymbol{0}$. The above equation simplifies to: 

\begin{align}
    \eta^{-1} = \frac{-1}{U}\int_{-0.5}^{0.5}{\boldsymbol{f}\cdot\boldsymbol{c} \, ds} = \frac{1}{U}\int_{-0.5}^{0.5}{\cos{\psi}\, ds} \label{eq:CostOfTransport}
\end{align}

The friction ratio $\mu$ does not appear explicitly in the cost of transport equation \ref{eq:CostOfTransport}. However, $\mu$ directly affects the velocity $U$, which is present in this equation. It is worth noting that $-\boldsymbol{f}\cdot\boldsymbol{c} = \cos(\psi)$ is the tangential friction force, and therefore according to equation \ref{eq:CostOfTransport}, the cost of transport can be determined as the ratio of the power consumed due to tangential friction ($\int_{-0.5}^{0.5}{-\boldsymbol{f}\cdot\boldsymbol{c} \, ds}$)  to the velocity of the snake $U$.

Snakes do not always optimize their energy consumption. To escape from a predator or to hunt, they may maximize their locomotion velocity $U$, which is determined in equation \ref{eq:U_equation}. Using the analytical framework for studying snake locomotion developed here, we aim to determine the optimal conditions for a body shape $\theta(s)$, for any given friction anisotropy ($\mu$)—that maximize (1) the overall efficiency $\eta$ and (2)  the total speed of the snake $U$.

Here, we employ \textit{variational methods} to address both the speed optimization and energy efficiency problems.
The shape and kinematics of the snake are described by a periodic shape function $\theta(s)$. The only constraint on this function is that it must ensure the continuity of the snake's body shape. To satisfy this constraint, we require $\int_{-0.5}^{0.5}{\sin\theta \, ds} = 0$. Note that smoothness or continuity of $\theta(s)$ itself is not necessary.  Since the system is kinematically compatible, for any given $\theta(s)$ function that satisfies this condition, there must be a corresponding solution for the system. This implies for any given $\theta(s)$, the snake will move with a velocity vector $\boldsymbol{U}$. This velocity vector can be determined using Newton's equation \ref{eq:NewtonsEquation}. However, we do not know the explicit form of $\boldsymbol{U}$ in terms of $\theta(s)$; we only know of its existence.  All kinematic variables, including $v(s)$, $\psi(s)$, $\boldsymbol{W}$, and $\boldsymbol{U}$ are functions of $\theta(s)$.

\subsection{Optimal Efficiency (\texorpdfstring{$\eta$}{eta})}

$\psi(s)$ and $\boldsymbol{U}$ are functions of $\theta(s)$. Therefore, equation \ref{eq:CostOfTransport} has the form of 
$\eta^{-1} = \int_{-0.5}^{0.5}{L(\theta)\, ds}$, where $L = \frac{\cos \psi}{U}$. Because $L$ is only a function of $\theta(s)$ and there is no $\theta'$, by substituting $L(\theta)$ into the Euler-Lagrange equation we obtain:

\begin{equation}
-\frac{d}{ds} \left( \frac{\partial L}{\partial \theta'} \right) + \frac{\partial}{\partial \theta} \left( L \right) =\frac{\partial}{\partial \theta} \left( \frac{\cos \psi}{U} \right) = 0,
\label{eq:eff_derivative_theta}
\end{equation}

and this could be correct only if $\frac{\cos\psi}{U}$ is constant. Because $\psi(s)$ and $U$ are only dependent on $\theta$. $U$ is independent of $s$ and therefore $\cos\psi$ must be constant at all body points. Considering equation \ref{eq:LocalVel} this leads to constant local velocity $v(s)$ and constant $\cos(\theta+\theta_0)$ for all $s$. The integral of the friction force in the y-direction over one cycle is zero: $\int_{-0.5}^{0.5}f_y \ ds = \int_{-0.5}^{0.5}\left(\cos\psi \sin\theta + \mu \sin\psi \cos \theta \right) ds = 0$. The first part of the integral is zero because $\cos\psi$ is constant and the integral of $\sin\theta$ is zero. Using $\sin\psi = \frac{w}{v} \sin(\theta+\theta_0)$ (equation \ref{eq:psi_theta_1}) and knowing that the integral of $\cos(\theta+\theta_0) \sin\theta$ over one cycle is zero, the above integral simplifies to $\int_{-0.5}^{0.5}f_y \ ds = \frac{\mu W}{v}\int_{-0.5}^{0.5} \sin\theta_0 \ ds$. This integral is zero only if $\theta_0 = 0$. Therefore, $\cos(\theta+\theta_0) = \cos{\theta}$ is constant if only $\theta = \pm constant$. A sawtooth shape is the most natural shape that satisfies this condition, but it is not the only solution. Any periodic continuous function composed of lines with angles of $\theta = \pm constant$ is a solution. also qualifies as a solution. The optimal shape belongs to this class of functions. However, we still need to determine the value of $\theta$ that minimizes the cost of transport.

Considering $\cos\theta$ is constant we conclude that $\lambda = \cos{\theta}$. Considering both $\cos{\theta}$ and $\cos{\psi}$ are constant, we found that the friction force in the $x$ direction is constant everywhere, which means it must be equal to zero, i.e. $f_x = \cos{\psi} \cos{\theta} - \mu \sin{\psi} \sin{\theta} = 0$. By combining this equation with equations \ref{eq:psi_theta_1}, \ref{eq:psi_theta_2}, and \ref{eq:LocalVel} and after replacing $\theta_0 = 0$, we can express the total velocity and cost of transport in terms of only the wavelength and friction ratio as follows:

\begin{align}
U &= \cos\theta-\frac{\cos\theta}{\cos^2\theta+\mu\sin^2\theta} = \lambda - \frac{\lambda}{\lambda^2+\mu (1-\lambda^2)} \label{eq:U_vs_theta} \\
\eta^{-1} &= \frac{1+(U-\lambda)\lambda}{U\sqrt{1+(U-\lambda)^2 + 2(U-\lambda)\lambda}}
\end{align}

Given a constant friction ratio $\mu$, we have derived expressions for speed ($U$) and efficiency ($\eta$) as functions of a single variable $\lambda$, where $0 < \lambda < 1$. To optimize efficiency, we perform single-variable optimization by setting the derivative of $\eta$ with respect to $\lambda$ to zero. Taking the total derivative of the efficiency with respect to wavelength ($\frac{d\eta}{d\lambda} = 0$) and simplifying it leads to a condition for optimal efficiency, which is expressed as the following sixth-order polynomial equation in $\lambda$:

\begin{align}
\left(\mu -1\right)^{2}\left(\mu+1\right)\lambda^{6}-3\mu\left(\mu-1\right)\left(\mu+1\right)\lambda^{4} + \mu\left(3\mu-2\right)\left(\mu+1\right)\lambda^{2}-\mu^{3}=0
\end{align}

\begin{figure}[ht]
  \centering
  \includegraphics[width=\textwidth]{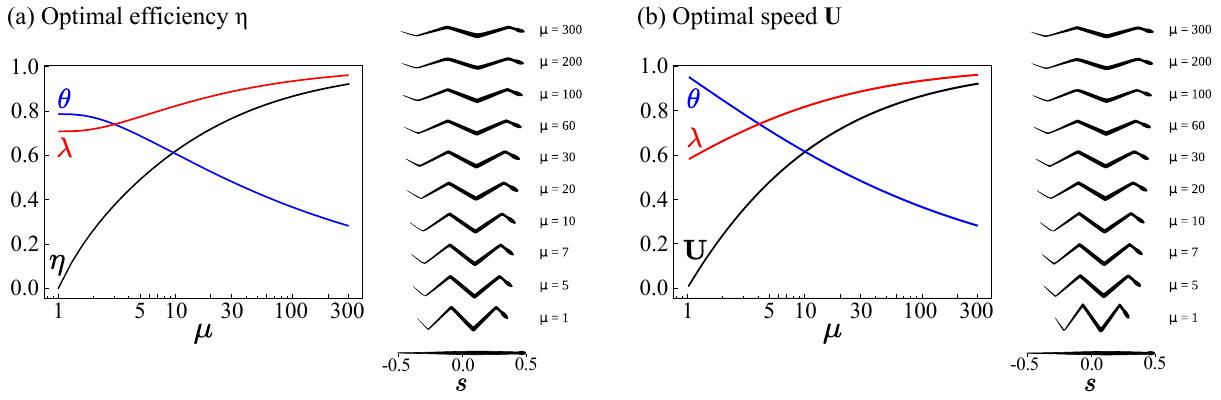}
  \caption{(a) Maximum efficiency $\eta$ and (b) maximum velocity $U$ for different friction ratio ($\mu$), and the corresponding wave length ($\lambda$) and body angle ($\theta$) for each case. The angles are in radian}
    \label{fig:UniformNormalResults}
\end{figure}

After solving the polynomial equation numerically, we identify the real roots of $\lambda$ that lie within the valid bounds, $0 < \lambda < 1$. To ensure the optimality of these solutions, we evaluate the second derivative of efficiency, $\frac{d^2\eta}{d\lambda^2}$, at each root. Only roots that satisfy the condition $\frac{d^2\eta}{d\lambda^2} < 0$ are retained, as they indicate local maxima of efficiency. For each friction ratio $\mu$, the maximum efficiency values, along with the associated $\lambda$ and angle $\theta$, are computed. Figure \ref{fig:UniformNormalResults}(a) illustrates how maximum efficiency varies with $\mu$, as well as the snake body shapes derived from these optimal values.

\subsection{Optimal Speed (\texorpdfstring{$U$}{U})} \label{sec:optimal_speed}

The same method can be used to find the maximum velocity. The total velocity is the average of the local velocity of the body points over one cycle. The magnitude of $U$ is expressed in integral form in equation \ref{eq:U_equation}. As explained in the previous section, once a periodic function for $\theta(s)$ is chosen, the other kinematic parameters, such as $v$, $\psi$, $\boldsymbol{W}$, and $\phi$, can be determined using the equilibrium equations. Thus, all the kinematic variables are functions of $\theta(s)$. However, we do not have explicit forms for these functions. To find the maximum $U$, we use the Euler-Lagrange equations and apply them to equation \ref{eq:U_equation}.

Everything inside the integral in this equation is a function of $\theta(s)$, and there is no $\theta' = \frac{\partial\theta}{\partial s}$. As a result, the Euler-Lagrange equation simplifies to:
\begin{equation}
\frac{\partial}{\partial \theta} \left( v \cos(\theta+\psi-\phi) \right) = 0,
\label{eq:U_derivative_theta}
\end{equation}
which implies that $v \cos(\theta+\psi-\phi)$ is constant. Referring to figure \ref{fig:RFTSchematic}(b), we can demonstrate that:
\begin{equation}
 v \cos(\theta+\psi-\phi) = -W\cos(\theta_0+\phi)+\cos(\theta-\phi),
\label{eq:v_in_U_direction}
\end{equation}
Since $W$, $\theta_0$, and $\phi$ are not functions of $s$, this expression can only remain constant if $\cos(\theta-\phi)$ is also constant. Similar to the previous section, the equilibrium equation in the $y$-direction can only be satisfied if $\phi = 0$ and $\theta = \pm constant$. Similar to achieving maximum efficiency, for maximum velocity, the snake should adopt a sawtooth shape. However, the angles may differ between these two scenarios. Since $\theta = \pm constant$, equation \ref{eq:U_vs_theta} is applicable to this case as well. Determining $\lambda$ to maximize $U$ in equation \ref{eq:U_vs_theta} can be approached either numerically or analytically. By taking the derivative of equation \ref{eq:U_vs_theta} with respect to $\lambda$ and setting it equal to zero, we can find the optimal 
$\lambda$ for any given $\mu$ to maximize $U$. The expression for the wavelength at maximum total speed is determined to be:

\begin{equation}
\lambda = \sqrt{\frac{2\mu+1 - \sqrt{8\mu +1}}{2(\mu - 1)}}
\end{equation}  

Figure \ref{fig:UniformNormalResults}(b) shows the results of speed optimization, indicating shapes that satisfy the optimal condition. These results are plotted to illustrate how optimized quantities such as $\lambda$ and $\theta$ vary with the friction coefficient ratio $\mu$.

\section{Discussion}

The systematic optimization of snake locomotion has been previously addressed in the work of Silas Alben \cite{doi:10.1098/rspa.2013.0236}, where a numerical approach was employed to identify optimal motions for locomotory efficiency across a broad range of frictional parameters. 
In his work the curvature $\frac{d \theta}{d s}$ of the body is parameterized using Chebyshev polynomials, providing a flexible representation of the snake body shapes. Through numerical optimization of the cost function, Alben explored a range of locomotion strategies and identified optimal retrograde traveling waves for a wide range of friction ratios, while triangular direct waves emerged as optimal in the limiting case of zero transverse friction.

In the present study, we developed an analytical framework to investigate the optimal locomotion of snakes. Assuming a constant average velocity for the snake and neglecting body rigidity, we demonstrated that both the energy efficiency and velocity can be expressed as functionals of the body shape, $\theta(s)$, as given in equations \ref{eq:CostOfTransport} and \ref{eq:U_equation}. Since the derivative of the shape function, $\theta'$, does not appear in these functionals, the application of the Euler-Lagrange equation to find the extrema results in simple equations. We found that for both optimal energy efficiency and maximum speed, the body should adopt a sawtooth shape, where $\theta = \pm constant$. For smaller friction ratios ($\mu<3$) the maximum energy efficiency is achieved when $\theta \approx 45^{\circ}$. However, achieving maximum velocity requires a steeper body angle, with lower friction ratios necessitating larger values of $\theta$, as evident in figure \ref{fig:UniformNormalResults}. For higher values of $\mu$, the body angles for both maximum velocity and maximum energy efficiency converge and the optimum $\theta$ decreases as
$\mu$ increases.

Further analysis reveals that at higher values of the friction ratio $\mu$, the behavior of the wavelength $\lambda$ in both optimal efficiency and speed converges to the same trend. Exploiting the closed-form solution for optimal speed derived in this study, we obtained an expression for the wave amplitude and its asymptotic behavior:

\begin{equation}
    A = \frac{1}{2}\sin{\theta} = \frac{1}{2}\sqrt{1 - \lambda^2} = \frac{1}{2}\sqrt{\frac{\sqrt{8\mu + 1} - 3}{2(\mu - 1)}} \sim \frac{1}{2}\left(\frac{\mu}{2}\right)^{-1/4}.
\end{equation}

\begin{figure}[ht]
    \centering
    \includegraphics[width=0.95\linewidth]{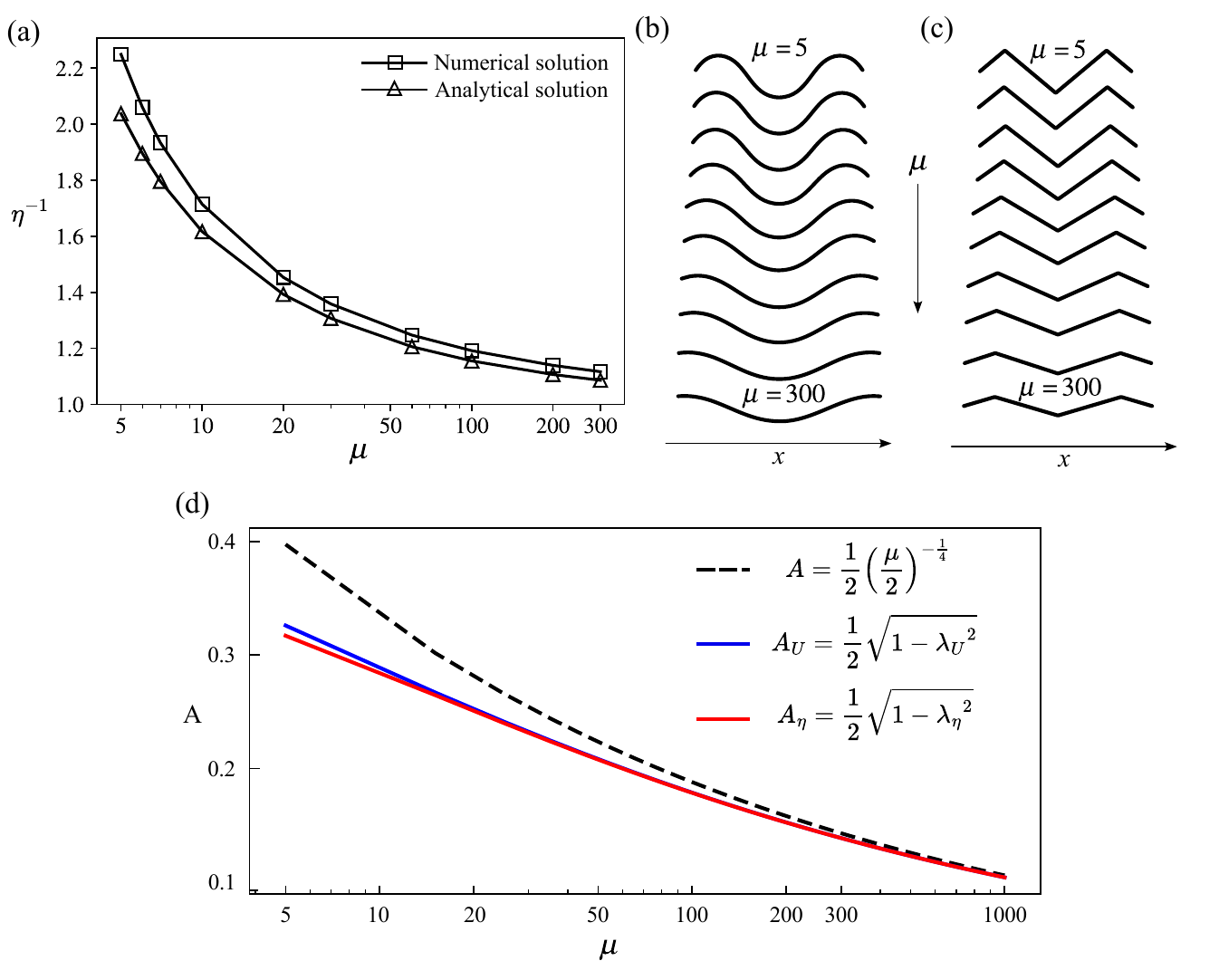}
    \caption{Comparison of optimization results and wave shapes from numerical and analytical methods: (a) cost of transport values versus friction coefficient ratio ($\mu$) obtained from numerical optimization (reconstructed from \cite{doi:10.1098/rspa.2013.0236}) and analytical optimization in this study. (b) Snake body shapes optimized using numerical methods for different $\mu$ values. (c) Snake body shapes optimized for efficiency using the analytical method proposed in this study. (d) Convergence of wave amplitudes ($A$) corresponding to $\lambda_U$ for optimal speed and $\lambda_\eta$ for optimal efficiency.}
    \label{fig:enter-label}
\end{figure}

This result aligns perfectly with the previous asymptotic analysis based on numerical optimization \cite{doi:10.1098/rspa.2013.0236}. The results of the numerical optimization and the proposed analytical framework for various values of $\mu$ are closely aligned. However, the analytical approach more effectively identifies body shapes that result in a lower cost of transport, i.e., higher efficiency (Figure \ref{fig:enter-label}). 

The method developed in this work can be applied to analyze locomotion in other media where the interactive forces can be described using resistive force theory (RFT). For example, the interaction between the body of an animal, such as a sandfish skink, and sand can be modeled using RFT \cite{doi:10.1126/science.1172490}. The forces in the normal and tangential directions between a moving element and sand are approximated as \cite{doi:10.1126/science.1172490}:

\begin{align}
f_n &= C_S \sin{\beta_0} + C_F \sin{\psi}
\\
f_t &= C_F \cos{\psi}
\end{align}

where $\tan\beta_0 = \cot{\gamma_0} \, \sin{\psi}$, and the constants $C_S$, $C_F$, and $\gamma_0$ characteristic parameters of the material response to drag. 
The cost of transport for the swimming of an elongated body in sand is analogous to snake locomotion on a flat surface. According to equation \ref{eq:CostOfTransport}, it can be determined as $\eta^{-1} = \frac{C_F}{U}\int_{-0.5}^{0.5}{\cos{\psi}\, ds}$. By applying the Euler-Lagrange equation for this case and following the same steps as for snake locomotion, we conclude that $\cos\psi$, $v$, and ultimately $\cos\theta$ must remain constant to achieve optimal locomotion. Consequently, a sawtooth body shape would also be ideal for swimming in sand. The angle of the sawtooth shape can be determined using Newton's equation \ref{eq:NewtonsEquation}.

Another example is the swimming of an elongated body such as flagellum in viscous fluids. The drag forces in the tangential and normal directions on an elongated segment in a low Reynolds number flow are proportional to the respective components of the relative velocity of the flow with respect to the body in those directions\cite{doi:10.1137/1.9781611970517}:

\begin{align}
f_n &= K_N \, v \sin{\psi}
\\
f_t &= K_T \, v \cos{\psi}
\end{align}

where $K_N$ and $K_T$ are constant numbers and for elongated bodies $K_N/K_T$ approaches to 2. These drag forces here are analogous to the friction forces experienced by snakes. However, unlike in snakes, the drag forces in a viscous fluid are proportional to velocity. The cost of transport for this case can be determined as $\eta^{-1} = \frac{K_T}{U}\int_{-0.5}^{0.5}{v \cos{\psi}\, ds}$. Therefore, using the Euler-Lagrange equation, $v \cos{\psi}$ must remain constant. Considering that $v(s)$ is merely a function of $\cos\psi$, we can conclude that $\cos\psi$ must be constant to achieve the extremum point. which leads to a sawtooth shape as in previous cases. Using the same method, we find that maximum velocity can also be achieved with a sawtooth shape, both for swimming in sand and viscous fluids. These findings are potentially applicable to undulatory locomotion in any medium where the interactive forces adhere to resistive force theory.

\section{Conclusion}
Lighthill \cite{doi:10.1137/1.9781611970517} previously demonstrated that the optimal kinematics for minimizing energy consumption in flagellar propulsion is a sawtooth shape. In this study, we used variational methods to construct an analytical framework for analyzing undulatory locomotion in any medium where forces can be determined using resistive force theory (RFT). We showed that the optimal kinematics for minimizing energy consumption or maximizing velocity in snake locomotion on flat surfaces, as well as in the swimming of an elongated body in sand or viscous fluids, is also a sawtooth shape. The angle of the sawtooth shape can be determined from Newton’s equations and depends on the characteristics of the interactive forces.

This framework has the potential to analytically explore more complex motions, such as 3D locomotion of snakes, where active regulation of weight distribution plays a crucial role in locomotion \cite{doi:10.1073/pnas.0812533106,hu2012slithering, Zhang2021}. In this work, we assumed negligible body rigidity. This raises an open question regarding the optimal kinematics when the energy consumption for body deformation is also taken into account. Another key assumption in this work, which is also made by Lighthill \cite{doi:10.1137/1.9781611970517}, is that the average velocity is constant, and there is no rotation during locomotion. However, for snakes with finite body length and a limited number of undulation cycles, the average velocity may periodically vary over each cycle, and there may also be a net angular displacement after each cycle. The results of the analytical framework presented here potentially provide an upper bound for energy efficiency and wave efficiency (speed) in more realistic scenarios, where snakes have finite lengths and experience internal energy dissipation due to body deformation. 

\bibliographystyle{unsrt}  
\bibliography{references}

\end{document}